\documentclass[12pt,dvips]{article}

\usepackage{rotating}
\usepackage{epsfig}
\usepackage{psfig}
\usepackage{color}

\newcommand{\lsim}{\raisebox{-0.13cm}{~\shortstack{$<$ \\[-0.07cm] $\sim$}}~}

\newcommand{\s}{\\ \vspace*{-3.5mm} }

\newcommand{\imag}{\Im {\rm m}}
\newcommand{\real}{\Re {\rm e}}

\begin{document}

\mbox{ } \\[-1cm]
\mbox{ }\hfill KAIST--TH--2002/14\\
\mbox{ }\hfill KEK--TH--826\\
\mbox{ }\hfill hep--ph/0206025\\
\mbox{ }\hfill \today\\
\bigskip

\thispagestyle{empty}
\setcounter{page}{0}

\begin{center}
{\Large{\bf CP--violating Chargino Contributions to the Higgs \\[0.5mm]
            Coupling to Photon Pairs in the Decoupling \\[2mm]
	    Regime of Higgs Sector}} \\[1.5cm]
  S.Y. Choi$^1$\footnote{E--mail address: sychoi@jinsol.chonbuk.ac.kr},\, 
  Byungchul Chung$^2$\footnote{E--mail address: crash@muon.kaist.ac.kr},\,
  P. Ko$^2$\footnote{E--mail address: pko@muon.kaist.ac.kr} and 
  Jae Sik Lee$^3$\footnote{E--mail address: jslee@post.kek.jp} 
\end{center}

\vskip 0.5cm

{\small
\begin{enumerate}
\item[{}] { }\ \ \ $^1${\it Department of Physics, Chonbuk National University,
            Chonju 561--756, Korea}\\[-7mm]
\item[{}] { }\ \  \ $^2${\it Department of Physics, KAIST, Daejeon 305--701, 
            Korea}\\[-7mm]
\item[{}] { }\ \  \ $^3${\it Theory Group, KEK, Tsukuba, Ibaraki 305--0801,
            Japan}
\end{enumerate}
}
\bigskip
\bigskip
\bigskip

\begin{abstract}
\noindent
In most supersymmetric theories, charginos $\tilde{\chi}^\pm_{1,2}$ 
belong to the class of the lightest supersymmetric particles and
the couplings of Higgs bosons to charginos are in general complex 
so that the CP--violating chargino contributions to the loop--induced 
coupling of the lightest Higgs boson to photon pairs can be sizable 
even in the decoupling limit of large pseudoscalar mass $m_A$
with only the lightest Higgs boson kinematically accessible at future
high energy colliders. We introduce a specific benchmark scenario of CP 
violation consistent with the electric dipole moment constraints
and with a commonly accepted baryogenesis mechanism in
the minimal supersymmetric Standard Model. Based on the benchmark scenario
of CP violation, we demonstrate that the fusion of the lightest Higgs boson 
in linearly polarized photon--photon collisions can allow us to confirm the
existence of the CP--violating chargino contributions {\it even in the 
decoupling regime of the Higgs sector} for nearly degenerate SU(2) gaugino 
and higgsino mass parameters of about the electroweak scale.
\end{abstract}
%


\newpage

\setcounter{footnote}{0}

\section{Introduction}

In the minimal supersymmetric extension of the Standard Model (MSSM)
\cite{MSSM}, the electroweak gauge symmetry is broken with two Higgs 
doublets, leading to the five physical states: a lighter CP--even Higgs 
boson $h$, a heavier CP--even Higgs boson $H$, a CP--odd Higgs boson $A$ 
and two charged Higgs bosons $H^\pm$ \cite{Higgs}. 
In some regions of MSSM parameter space, more than one Higgs boson
can be discovered at the LHC. However, there exists a sizable region of the
parameter space at moderate values of the ratio of two Higgs vacuum expectation
values, $\tan\beta=v_2/v_1$, opening up from about $m_A=200$ GeV to higher 
values in which the heavier Higgs bosons cannot be discovered at the 
LHC \cite{LHC}. 
In this so--called decoupling regime of large $m_A$, only the lightest 
Higgs boson can be discovered not only at
the LHC but also at the first stage of $e^+e^-$ linear colliders (LC)
\cite{LC1} and the properties of the lightest Higgs boson are nearly 
indistinguishable from those of the Standard Model (SM) Higgs boson. 
Therefore, high precision measurements of branching ratios and 
other properties of the lightest Higgs boson \cite{LC2,DKZ,DDHI} are required 
in order to detect any deviations from SM Higgs predictions.\s

Deviations from SM properties in the decoupling limit can occur if 
the Higgs decay into supersymmetric particles are kinematically 
allowed \cite{DKZ}, or if light supersymmetric particles contribute 
significantly to Higgs loop amplitudes \cite{DDHI}. In this light, the 
colliding $\gamma$ beam reaction
\begin{eqnarray}
\gamma\gamma\rightarrow h\,,
\label{eq:rrh}
\end{eqnarray}
has been regarded as an important mechanism \cite{GGH,GG,KKSZ,CL1,BCK}
for probing the properties of the Higgs boson $h$ precisely. 
Since the coupling of $h$ to photon pairs is mediated by loops of all 
charged particles with non--zero mass, the measurements of the coupling 
through the process (\ref{eq:rrh}) can virtually reveal the action of 
new particle species and help discriminate the lightest SUSY Higgs boson 
from the SM Higgs boson even in the decoupling regime of large $m_A$. 
In the decoupling regime, the chargino contributions to the $h\gamma\gamma$ 
coupling are responsible for most of the deviation from the SM for  
low $\tan\beta$ and nearly degenerate SU(2) gaugino and higgsino mass 
parameters $M_2$ and $\mu$ of the electroweak scale, and
the top squark contributions to the $h\gamma\gamma$ coupling can be in general
sizable for light top squark masses, in particular, in the maximal mixing 
scenario with large top squark mixing parameter \cite{DDHI}.\s

In CP--noninvariant SUSY theories, the spin--1/2 charginos unlike any 
spin--zero charged particles have a distinct feature; while spin--zero top 
squarks contribute only to the CP--even part of the $h\gamma\gamma$ coupling, 
charginos can contribute to the CP--odd as well as the CP--even 
parts \cite{BCK}, leading to CP violation. This possible CP--violating 
coupling of the Higgs boson $h$ to photon pairs can be directly probed 
through the process $\gamma\gamma\rightarrow h$ by using high energy 
colliding beams of linearly polarized photons \cite{GG,KKSZ,CL1,BCK},  
generated by Compton back--scattering of linearly polarized laser light on 
electron/positron bunches at a LC \cite{GKPST}.\s

The prime goal of the present note is to probe the possibility of measuring 
the CP--violating chargino contributions to the loop--induced $h\gamma\gamma$ 
coupling {\it in the decoupling regime of large $m_A$} in detail through 
the process $\gamma\gamma\rightarrow h$ with linearly polarized 
photon--photon collisions, based on a specific benchmark scenario of CP 
violation consistent with all the present indirect constraints and direct
searches.\s

The paper is organized as follows. In the next section, the chargino mixing 
and the couplings of the lightest Higgs boson to diagonal chargino pairs 
are described; the sum rules relating two couplings also are presented. 
Based on the recent works \cite{BAU} for electroweak baryogenesis in the 
framework of MSSM and the present experimental constraints on the lightest 
Higgs boson mass and the lighter chargino mass \cite{MCX} as well as
the electric dipole moment (EDM) constraints \cite{EDM,TWO1,TWO2}, we introduce 
a feasible benchmark scenario with maximal CP--violating phase 
in the chargino mass matrix. Section~\ref{sec:production} is devoted to 
a brief review of the production of the lightest Higgs boson through 
collisions of two back--scattered photons and of the polarization 
asymmetries for probing CP violation. After discussing the CP--violating 
chargino contributions to the $h\gamma\gamma$ coupling in 
Sect.~\ref{sec:charginos}, we present a detailed numerical study of the 
CP--violating chargino as well as right--handed top squark contributions to 
the coupling of the lightest Higgs bosons to photon pairs in
Sect.~\ref{sec:numerical}, based on the benchmark scenario of CP violation.
Conclusions are finally given in Section~\ref{sec:conclusion}.\s

\section{The lightest Higgs boson couplings to chargino pairs}
\label{sec:coupling}

The chargino masses and their couplings to Higgs particles 
are determined by the SU(2) gaugino mass parameter $M_2$ and the
higgsino mass parameter $\mu$. In standard definition 
\cite{CKMZ}, the diagonalization $U_R\,{\cal M}_C\, U^\dagger_L
={\rm diag}\{m_{\tilde{\chi}^\pm_1}, m_{\tilde{\chi}^\pm_2}\}$ of the 
chargino matrix ${\cal M}_C$ by two unitary matrices $U_L$ and $U_R$ 
in the MSSM 
\begin{eqnarray}
{\cal M}_C=\left(\begin{array}{cc}
                M_2                &  \sqrt{2}m_W c_\beta \\[2mm]
             \sqrt{2}m_W s_\beta  &             \mu   
                  \end{array}\right)\,,
\label{eq:mass matrix}
\end{eqnarray}
generates the light and heavy chargino states $\tilde{\chi}^\pm_i$ ($i=1,2$), 
ordered with rising mass. The coefficients $s_\beta =\sin\beta$, 
$c_\beta=\cos\beta$ are given by $\tan\beta$, and $s_W=\sin\theta_W,\, 
c_W=\cos\theta_W$ are the sine and cosine of the electroweak mixing angle. 
In CP--noninvariant theories, the mass parameters are complex.\s

By reparametrization of the field basis, the SU(2) mass parameter $M_2$ in 
Eq.(\ref{eq:mass matrix}) can always be set real and positive, while  
the higgsino mass parameter $\mu$ is assigned the phase $\Phi_\mu$. 
The chargino masses are then given in terms of the real and positive
parameters $\tan\beta$, $M_2$ and $|\mu|$ and the phase $\Phi_\mu$ by
\begin{eqnarray}
m^2_{\tilde{\chi}^\pm_{1,2}} =\frac{1}{2}\left[\, M^2_2+|\mu|^2+2\, m^2_W\mp
 \sqrt{(M^2_2+|\mu|^2+2 m^2_W)^2-4|m^2_W\,s_{2\beta}
        -M_2 |\mu|\, {\rm e}^{i\Phi_\mu}|^2}\,\right],
\label{eq:masses}
\end{eqnarray}
which is symmetric with respect to the parameters $M_2$ and $\mu$. Here,
$s_{2\beta}=\sin 2\beta$. The coupling of the lightest Higgs boson to 
diagonal chargino pairs in the decoupling regime
is given in terms of the parameter $\tan\beta$ and the matrices $U_{L,R}$ by 
\begin{eqnarray}
\langle\tilde{\chi}^-_{iR}|h|\tilde{\chi}^-_{iL}\rangle \equiv \kappa_i
  = -\frac{g}{\sqrt{2}}\left(U_{Ri1} U^*_{Li2}\, c_\beta 
                            +U_{Ri2} U^*_{Li1}\, s_\beta
		       \right)\,,
\label{eq:hcc}
\end{eqnarray}
and $\langle\tilde{\chi}^-_{iL}|h|\tilde{\chi}^-_{iR}\rangle =
\langle\tilde{\chi}^-_{iR}|h|\tilde{\chi}^-_{iL}\rangle^* = \kappa^*_i$.
We note that the diagonal couplings $\kappa_i$ depend symmetrically on 
$M_2$ and $\mu$ because the interchange of $M_2$ and $\mu$ in the chargino 
mass matrix ${\cal M}_C$ is compensated simply by the interchanges,
$U_{Ri1}\leftrightarrow U^*_{Li2}$ and $U_{Ri2}\leftrightarrow U^*_{Li1}$,
giving the same chargino mass eigenvalues and leaving $\kappa_i$ 
invariant.  Moreover, with the off--diagonal 
entries, $\sqrt{2}m_W c_\beta$ and $\sqrt{2} m_W s_\beta$, 
of the chargino mass matrix ${\cal M}_C$,
the couplings of the lightest Higgs boson to chargino pairs, $\kappa_i$, and
the chargino masses $m_{\tilde{\chi}^\pm_{1,2}}$ satisfy the sum rules:
\begin{eqnarray}
&&   m_{\tilde{\chi}^\pm_1}\,\real(\kappa_1)\, 
   + m_{\tilde{\chi}^\pm_2}\,\real(\kappa_2)\, 
   = - g m_W\,,\nonumber\\[2mm]
&&   m_{\tilde{\chi}^\pm_1}\imag(\kappa_1)
   + m_{\tilde{\chi}^\pm_2}\imag(\kappa_2) = 0\,.
\label{eq:sum rules}
\end{eqnarray}
As a result, it is easy to see that in case of degenerate chargino masses, 
perfect cancellation occurs, causing no CP violation. In addition,
the condition for the vanishing scalar coupling of the 
neutral Goldstone boson $G^0$ to chargino pairs leads to an
interesting relation, $s_\beta\, \imag(U_{Ri2}U^*_{Li1})= c_\beta\, 
\imag(U_{Ri1}U^*_{Li2})$, forcing $\imag (k_i)$ to be 
proportional to $s_{2\beta} \sin\Phi_\mu$ \cite{TWO2}.\s

The couplings (\ref{eq:hcc}) clearly show that the Higgs boson
$h$ couples to mixtures of gaugino and higgsino components of charginos.
In particular, if the light chargino $\tilde{\chi}^\pm_1$ were 
either a pure gaugino ($|\mu|\gg |M_2|$) or a pure higgsino state 
($|M_2|\gg |\mu|$), the $h\tilde{\chi}^+_i\tilde{\chi}^-_i$ couplings vanish. 
Only when the higgsino parameter $\mu$ is comparable to the gaugino mass 
parameters $M_2$ in size and the value of $\tan\beta$ is moderate, 
the coupling of the Higgs boson to diagonal chargino pairs 
could be significant. Consequently, the chargino contributions
to the $h\gamma\gamma$ coupling can be significant (only) for moderate 
$\tan\beta$ and almost degenerate $M_2$ and $|\mu|$ of the electroweak scale.\s

It is also worthwhile to note that the most likely scenario for generating 
enough baryon asymmetry of the universe through electroweak baryogenesis 
in the framework of MSSM is to make use of the phase 
$\Phi_\mu$ \cite{BAU}.  In this scenario the phase $\Phi_\mu$ should 
be close to maximal, and the left--handed top squark must be very heavy 
to give sufficiently large radiative corrections to the lightest Higgs 
boson mass, given the need for a light right--handed top squark to get 
a strongly first order electroweak 
phase transition with small $\tan\beta$ and large pseudoscalar mass $m_A$. 
Moreover, the MSSM chargino baryogenesis requires nearly degenerate
SU(2) gaugino and higgsino mass parameters $M_2$ and $|\mu|$ of the electroweak
scale.\s

\begin{figure}[htbp]
 \hspace*{0.0cm}
 \centerline{\psfig{figure=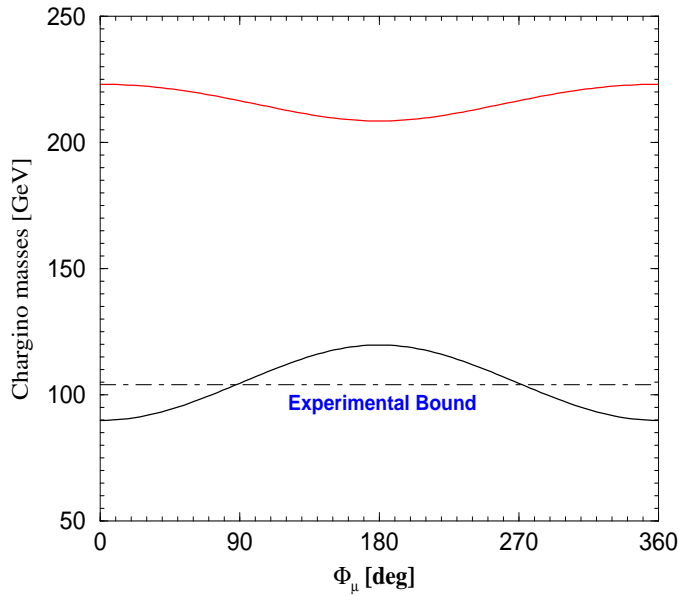,height=8cm,width=9cm}}
 \vspace*{-0.4cm}
 \caption{\it The chargino masses, $m_{\tilde{\chi}^\pm_{1,2}}$ as a function
           of the CP--violating phase $\Phi_\mu$; the other SUSY parameters are
 	  given as in Eq.(\ref{eq:MCPX}). The lower (upper) line is for 
 	  the lighter (heavier) chargino mass; the dot--dashed line indicates 
 	  the present experimental lower bound of about 104 GeV on the 
 	  lighter chargino mass \cite{MCX}.}
\label{fig:fig1}
\end{figure}

Motivated by the above observations for MSSM chargino baryogenesis 
and by the present experimental constraints on the lightest Higgs boson mass 
and lighter chargino mass \cite{MCX} as well as the experimental EDM
constraints \cite{EDM}, we consider
the following CP--violating benchmark scenario of the lightest Higgs boson
mass $m_h$ and the SUSY parameters:
\begin{eqnarray}
m_h=115\, {\rm GeV};\ \ \tan\beta=5; \ \ M_2=150\, {\rm GeV}, 
\ \ |\mu|=150\, {\rm GeV}; \ \ \Phi_\mu=\frac{\pi}{2}\,,
\label{eq:MCPX}
\end{eqnarray}
and two values of the right--handed top squark mass, $m_{\tilde{t}_R}=100$ and 
$250$ GeV.
For the parameter set (\ref{eq:MCPX}) the lighter chargino mass turns out to
be 105 GeV, slightly larger than the experimental lower bound of 
about 104 GeV set for the electron sneutrino mass exceeding 300 GeV and the
difference of the lighter chargino and lightest neutralino masses larger
than 4 GeV \cite{MCX}, which is supported by supergravity and gauge--mediation
models with the gaugino mass unification $|M_1|\simeq 0.5 M_2$. (In the
anomaly--mediation models with inverse gaugino mass hierarchy \cite{AMSB}, the
experimental chargino mass bound could be much weaker, because of
the almost degenerate lighter chargino and lightest neutralino states.)\,   
The lighter chargino mass
$m_{\tilde{\chi}^\pm_1}$ is larger for negative $\cos\Phi_\mu$ 
and 
for nearly degenerate $M_2$ and $|\mu|$
as can be checked in Eq.(\ref{eq:masses}) and from Fig.\ref{fig:fig1}. 
If $M_2$ and/or $|\mu|$ are much larger than $m_W$, the lighter chargino 
mass is then nearly equal to ${\rm min}\{M_2, |\mu|\}$ and
almost independent of $\tan\beta$ and $\Phi_\mu$.\s

\vskip 0.3cm

There may exist important constraints \cite{EDM} on the CP phase 
$\Phi_\mu$ in the MSSM from experimental limits on the EDMs of the electron, 
neutron and $^{199}{\rm Hg}$. However, the one--loop level 
contributions to the EDMs from sfermions, charginos and neutralinos in the 
MSSM are strongly suppressed if at least first and second generation sfermions 
are much heavier than sfermions of the third generation \cite{EDM}. 
Moreover, we find that the two--loop level contributions from top squarks 
and/or charginos \cite{TWO1,TWO2}, present even in case of very heavy 
first-- and second--generation sfermions, are significantly suppressed in the 
decoupling limit of large $m_A$. Consequently, we emphasize that the entire 
range between $0$ and $2\pi$ for the phase $\Phi_\mu$ can still be allowed 
in the decoupling limit of large pseudoscalar mass $m_A$ and for heavy 
sfermions of the first and second generations.\s

For the parameter set (\ref{eq:MCPX}) some chargino and neutralino states 
as well as the light right--handed top squark state are accessible 
experimentally at the first phase of the LC. Then, the measurement of the 
masses and production cross sections with polarized $e^+e^-$ beams will 
allow us to determine the parameters, $\{\tan\beta, M_2,\mu\}$ and the U(1) 
gaugino mass parameter $M_1$ \cite{LC2,DKZ} as well as the top squark mass 
$m_{\tilde{t}_R}$ with good precision \cite{LC2}. 
Therefore, we can check whether the indirect measurements of the chargino 
and top squark systems through the process $\gamma\gamma\rightarrow h$ are 
consistent with their direct measurements through pair production in 
$e^+e^-$ collisions or not. \s

\section{The lightest Higgs boson production at $\gamma\gamma$ colliders}
\label{sec:production}

The helicity amplitude of the reaction $\gamma\gamma\rightarrow h$ in the 
two--photon center--of-mass frame can be in general written 
as
\begin{eqnarray}
{\cal M}_{\lambda_1\lambda_2} = - m_h \frac{\alpha}{4\pi}
  \left[\, S(m_h) + i \lambda_1 P(m_h)\right]\, 
  \delta_{\lambda_1\lambda_2}\,,
\label{eq:helicity amplitude}
\end{eqnarray}
with the photon helicities $\lambda_{1,2}=\pm 1$. Here, the form factor
$S(m_h)$ ($P(m_h)$) represents the CP--even (--odd) coupling strength of
the Higgs boson to two photons; the simultaneous existence of both the 
CP--even and CP--odd form factors signals CP violation. In addition to
the unpolarized cross section $\sigma_0 (s) \equiv \hat{\sigma}_0 \,
\delta(1-m^2_h/s)$ with 
\begin{eqnarray}
\hat{\sigma}_0 \equiv \frac{\pi}{4\, m^4_h}
           \left[\, |{\cal M}_{++}|^2 + |{\cal M}_{--}|^2\right]\,
	   = \frac{\alpha^2}{32\pi m^2_h}\left(|S|^2+|P|^2\right)\,,
\label{eq:cross section}
\end{eqnarray}
we can define three polarization asymmetries in terms of the helicity 
amplitude (\ref{eq:helicity amplitude}) as
\begin{eqnarray}
{\cal A}_1 &\equiv& \frac{|\,{\cal M}_{++}|^2-|\,{\cal M}_{--}|^2}{
                          |\,{\cal M}_{++}|^2+|\,{\cal M}_{--}|^2}
           \, =\, \frac{2\, \imag (S P^*)}{|S|^2+|P|^2}\,, \nonumber\\[1mm]
{\cal A}_2 &\equiv& \frac{2\, \imag (\,{\cal M}^*_{--} {\cal M}_{++})}{
                          |\,{\cal M}_{++}|^2+|\,{\cal M}_{--}|^2}
           \, =\, \frac{2\, \real (S P^*)}{|S|^2+|P|^2}\,, \nonumber\\[1mm]
{\cal A}_3 &\equiv& \frac{2\, \real (\,{\cal M}^*_{--} {\cal M}_{++})}{
                          |\,{\cal M}_{++}|^2+|\,{\cal M}_{--}|^2}
           \, =\,  \frac{|S|^2-|P|^2}{|S|^2+|P|^2} \,.
\end{eqnarray}
In CP--invariant theories, the form factors $S$ and $P$ cannot be
simultaneously non--vanishing, leading to the relations, ${\cal A}_1={\cal 
A}_2=0$ and ${\cal A}_3=\pm 1$ such that ${\cal A}_1\neq 0$, 
${\cal A}_2\neq 0$ or $|{\cal A}_3| \, <\, 1$ signals CP violation.
Note that the asymmetry ${\cal A}_1$ is non--zero only when $S$ and $P$
have a non--zero relative phase, which can be developed when
the lightest Higgs boson mass is larger than twice the mass of any charged 
particles in the loop. Neglecting the tiny contributions from light quarks 
to the $h\gamma\gamma$ coupling, one can safely assume that both $S$ and $P$ 
are real, because the lightest Higgs boson mass $m_h$ is less than twice the 
$W$ boson mass as well as twice the experimental lower bound of about 104 GeV 
on the lighter chargino mass.\s

After folding the luminosity spectra of two linearly polarized photon beams, 
the event rate of the Higgs boson production via two--photon fusion is given by
\begin{eqnarray}
N_h(P_T, \bar{P}_T, \eta) 
   = N_h^0\, \left[1 
   +P_T \bar{P}_T\left(\, {\cal A}_2 \sin 2\eta 
                         + {\cal A}_3 \cos 2\eta\right)
    \, \frac{\langle f_3 * f_3\rangle_\tau}{
             \langle f_0 * f_0\rangle_\tau}
   \right]\,,
\label{eq:polarized number}
\end{eqnarray}
with $\tau=m^2_h/s$ taken, with $\{P_T, \bar{P}_T\}$ the 
degrees of linear polarization of the initial laser photon beams, 
respectively, and with $\eta$ the azimuthal angle between the directions 
of maximal linear polarization of two initial laser beams. Here, the averaged
number of Higgs bosons $N_h^0$ is given by \cite{GG}
\begin{eqnarray}
N^0_h &\equiv& \hat{\sigma}_0\, 
               \left. s_{\gamma\gamma}
	       \frac{d L_{\gamma\gamma}}{d s_{\gamma\gamma}}
               \right|_{s_{\gamma\gamma}=m^2_h} 
	  = \frac{8\pi^2\Gamma(h\rightarrow \gamma\gamma)}{m^3_h}
               \left. s_{\gamma\gamma}
	       \frac{d L_{\gamma\gamma}}{d s_{\gamma\gamma}}
               \right|_{s_{\gamma\gamma}=m^2_h} 
	       \nonumber\\[1mm]
      &\simeq& 1.54\times 10^2 \left(\frac{L_{ee}}{\,100\,{\rm fb}^{-1}}\right)
         \left(\frac{{\rm TeV}}{\sqrt{s}}\right)
         \frac{\Gamma(h\rightarrow \gamma\gamma)}{{\rm keV}}
	 \left(\frac{{\rm 100\, GeV}}{{\rm m_h}}\right)^2 F(m_h)\,,
\label{eq:averaged number}
\end{eqnarray}
where $\Gamma(h\rightarrow\gamma\gamma)$ is the two--photon decay width of
the Higgs boson $h$, and $F(\sqrt{s}_{\gamma\gamma})=(\sqrt{s}/L_{ee})\,\, 
{d L_{\gamma\gamma}}/{d \sqrt{s}_{\gamma\gamma}}$ is a slowly varying function
dependent upon the details of the machine design, but it could
be of the order of unity. In Eq.(\ref{eq:averaged number}), $s$ 
($s_{\gamma\gamma}$) and $L_{ee}$ ($L_{\gamma\gamma}$) are the 
$e^+e^-$ $(\gamma\gamma)$ c.m. energy squared and integrated luminosity, 
respectively. The integrated luminosity $L_{ee}$ of about 1 ab$^{-1}$ is
expected to be accumulated within a few years in the first phase of a LC 
with a clean experimental environment \cite{LC1}. Numerically, 
$\Gamma(h\rightarrow\gamma\gamma)\simeq 4.4$ keV for the parameter 
set (\ref{eq:MCPX}) with $m_{\tilde{t}_R}=100$ GeV so that for 
$\sqrt{s}=500$ GeV or less the number of Higgs boson events $N^0_h$ is 
expected to be as large as 10000 with such a high integrated luminosity.\s

The ratio of the luminosity correlations, 
$\langle f_3 * f_3\rangle_\tau/\langle f_0 * f_0\rangle_\tau$ in
Eq.~(\ref{eq:polarized number}), whose 
definition can be found in Ref.\cite{KKSZ,CL1}, depends on the beam energy 
$\sqrt{s}$ and the laser frequency $\omega_0$. As studied in detail 
in Ref.\cite{KKSZ,CL1}, the ratio reaches its maximal value for small values 
of $x=2\sqrt{s} \, \omega_0/m^2_e$ and near the upper bound of 
$\tau\lsim \tau_{\rm max}= x^2/(1+x)^2$, i.e. if the energy is just 
sufficient to produce the lightest Higgs boson in the $\gamma\gamma$ 
collisions; since the luminosity vanishes at $\tau=\tau_{\rm max}$, the 
operating condition should be adjusted properly such that $\tau\lsim 
\tau_{\rm max}$ allows for a sufficiently large luminosity $\langle 
f_0 * f_0\rangle$. The coefficient of ${\cal A}_2$ (${\cal A}_3$)  
in Eq.(\ref{eq:polarized number}) is proportional to 
$P_T\bar{P}_T\, \sin 2\eta$ ($P_T\bar{P}_T\, \cos 2\eta$) so that
setting $P_T=1$ and $\bar{P}_T=1$, the CP--odd asymmetry ${\cal A}_2$ 
is separated by taking the difference of cross sections for the azimuthal 
angle $\eta=\pi/4$ and $-\pi/4$, while the CP--even asymmetry ${\cal A}_3$
would be determined by the difference of cross sections for $\eta=0$ and 
$\pi/2$.\s

\section{Chargino contributions to the $h\gamma\gamma$ coupling}
\label{sec:charginos}

Neglecting the contributions of the left--handed top squarks and other squarks,
(assumed to be very heavy in favor of electroweak baryogenesis and the EDM
constraints), as well as those from the light charged fermions, the CP--even 
form factor $S$ in the decoupling regime can be decomposed into the top quark, 
$W^\pm$ boson, right--handed top squark and chargino parts but the CP--odd 
form factor $P$ consists {\it only} of the chargino contributions: 
\begin{eqnarray}
&& S = S_t + S_{W^\pm}+S_{\tilde{t}_R}+S_{\tilde{\chi}^\pm_1}
      +S_{\tilde{\chi}^\pm_2}\,,\nonumber\\
&& P = P_{\tilde{\chi}^\pm_1}+ P_{\tilde{\chi}^\pm_2}\,.
\end{eqnarray}
The explicit form of the CP--even functions, $S_t$, $S_{W^\pm}$ and
$S_{\tilde{t}_R}$, can be found in Ref.\cite{CL1}. 
The SM contributions, $S_t$ and $S_{W^\pm}$, depend only on the Higgs boson 
mass $m_h$, while the right--handed top squark contribution on its mass
$m_{\tilde{t}_R}$ as well as the lightest Higgs boson mass and it
decreases very quickly $\propto 1/m^2_{\tilde{t}_R}$ with increasing
right--handed top squark mass. The CP--even and CP--odd amplitudes  
$S_{\tilde{\chi}^\pm_i}$ and $P_{\tilde{\chi}^\pm_i}$ ($i=1,2$) for the
chargino contributions are given in terms of the diagonalization matrices 
$U_{L,R}$ and the chargino masses $m_{\tilde{\chi}^\pm_{1,2}}$ by \cite{BCK}
\begin{eqnarray}
&& S_{\tilde{\chi}^\pm_i}= + 2 \real (\kappa_i) \,
       \frac{m_h\,\,}{m_{\tilde{\chi}^\pm_i}}\, F_{sf}(\tau_i)\,, \nonumber\\
&& P_{\tilde{\chi}^\pm_i}= -2 \imag (\kappa_i)\, 
       \frac{m_h\,\,}{m_{\tilde{\chi}^\pm_i}}\, F_{pf}(\tau_i)\,,
\end{eqnarray}
with $\tau_i=m^2_h/4m^2_{\tilde{\chi}^\pm_i}$ ($i=1,2$) and 
$\kappa_i=\langle \tilde{\chi}^-_{iR}|h|\tilde{\chi}^-_{iL}\rangle$ in 
Eq.(\ref{eq:hcc}). With the help of the so--called scaling function 
$f(\tau_i)$, which is ${\sf arcsin}^2(\sqrt{\tau_i})$ for $\tau_i\leq 1$, 
the functions $F_{sf}(\tau_i)$ and 
$F_{pf}(\tau_i)$ are given by  
\begin{eqnarray}
&& F_{sf} (\tau_i) = \tau^{-1}_i\left[1+(1-\tau^{-1}_i) 
                     f(\tau_i)\right]\,, \nonumber\\
&& F_{pf} (\tau_i) = \tau^{-1}_i f(\tau_i)\,.
\end{eqnarray}
In the limit of heavy loop masses ($\tau_i\rightarrow 0$), these amplitudes 
reach the asymptotic values, $F_{sf}\rightarrow 2/3$ and $F_{pf}\rightarrow 1$,
the minimum values in the range of $\tau_i\leq 1$.
Note that the contributions of the charginos vanish in the 
large chargino mass limit since the amplitudes $S_{\tilde{\chi}^\pm_i}$ and
$P_{\tilde{\chi}^\pm_i}$ are damped by the heavy chargino masses $\propto
1/m_{\tilde{\chi}^\pm_i}$.\s

\section{Numerical results}
\label{sec:numerical}

First of all, let us investigate the dependence of the amplitudes 
$S_{\tilde{\chi}^\pm_i}$ and $P_{\tilde{\chi}^\pm_i}$ on the relevant 
SUSY parameters $\{\tan\beta, M_2, |\mu|, \Phi_\mu\}$ from the standpoint of
the benchmark scenario (\ref{eq:MCPX}).  With the lighter chargino 
contribution larger than the heavier chargino contribution to the 
$h\gamma\gamma$ coupling and with the sum rules (\ref{eq:sum rules})
relating the two chargino contributions, it is sufficient to present the 
amplitudes $S_{\tilde{\chi}^\pm_1}$ and $P_{\tilde{\chi}^\pm_1}$. \s

\begin{figure}[htp]
\vskip 0.3cm
 \hspace*{0.0cm}
 \centerline{\psfig{figure=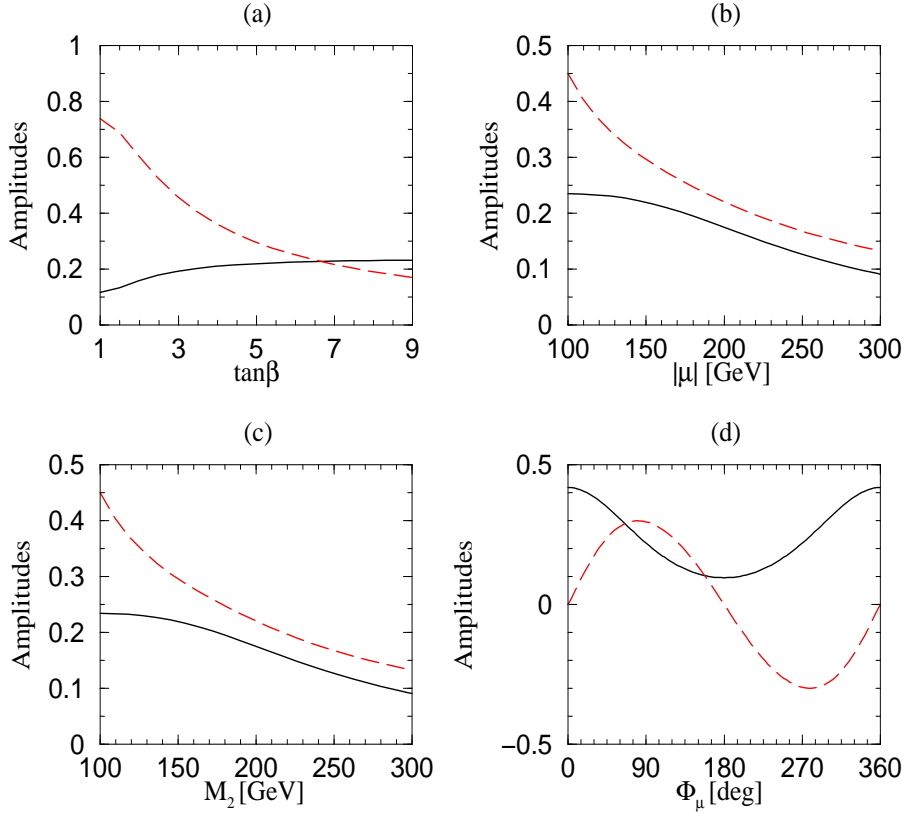,height=11cm,width=12cm}}
 \vspace*{-0.9cm}
 \caption{\it The CP--even and CP--odd amplitudes, $S_{\tilde{\chi}^\pm_1}$
          (solid lines) and $P_{\tilde{\chi}^\pm_1}$ (dashed lines), for the 
	  light chargino loop contribution as a function of (a)  
	  $\tan\beta$, (b) $|\mu|$, (c) $M_2$ and (d) the phase $\Phi_\mu$; 
	  in each figure, the other parameters except for each varied 
	  parameter are given as in Eq.(\ref{eq:MCPX}). Here, 
	  $m_{\tilde{t}_R}=100$ GeV is assumed. For comparison,
	  we note that the sum of the SM contributions, $S_t+S_{W^\pm} 
	  \simeq 2.9$. }
\label{fig:fig2}
\vskip 0.3cm
\end{figure}

Figure~\ref{fig:fig2} shows the dependence of the CP--even and CP--odd 
amplitudes, $S_{\tilde{\chi}^\pm_1}$ (solid lines) and 
$P_{\tilde{\chi}^\pm_1}$ (dashed lines) for the lighter chargino loop 
contribution on (a) $\tan\beta$, (b) $|\mu|$, (c) $M_2$ and (d) the phase 
$\Phi_\mu$, respectively. For the sake of comparison, we note that
the sum of the SM contributions, $S_t+S_{W^\pm}\simeq 2.9$ in size for 
$m_h=115$ GeV, $m_W=80$ GeV and $m_t=165$ GeV at the electroweak scale.
In each figure the other parameters except for 
each varied parameter are assumed to be given as in Eq.(\ref{eq:MCPX}) and
$m_{\tilde{t}_R}$ is taken to be 100 GeV.
Only the region with $|\mu|\geq 150$ GeV and  $M_2 \geq 150 $ GeV
in the frames (b) and (c), respectively, and that with $90^{\rm o}\leq
\Phi_\mu \leq 270^{\rm o}$ in the frame (d) satisfy the constraint 
$m_{\tilde{\chi}^\pm_1} \simeq 104$ GeV. 
Nevertheless, we show the amplitudes for all the ranges because
the experimental chargino mass bound is model--dependent \cite{AMSB}. \s
 
Firstly, we note that as $\tan\beta$ increases the CP--odd amplitude
$P_{\tilde{\chi}^\pm_1}$ decreases, but the CP--even amplitude
$S_{\tilde{\chi}^\pm_1}$ increases (see the upper--left frame of
Fig.\ref{fig:fig1}). This different behavior is due to
the fact that with increasing $\tan\beta$ the lighter chargino mass
$m_{\tilde{\chi}^\pm_1}$ decreases for $\Phi_\mu=\pi/2$, $\imag (\kappa_1)$ 
decreases in proportion to $s_{2\beta}= 2\tan\beta/(1+\tan^2\beta)$, and 
$\real(\kappa_1)$ approaches to a constant value. Secondly, as shown
in the upper--right and lower--left frames the dependence of 
both the CP--even and CP--odd amplitudes on $M_2$ and $|\mu|$ is
identical as expected from the identical dependence of $\kappa_1$ and 
the chargino masses on the parameters, and both amplitudes eventually 
decrease as either $M_2$ and $|\mu|$ increases. This reflects
that if either $M_2$ or $|\mu|$ is very large the lighter chargino state 
is almost higgsino--like or gaugino--like, respectively, and 
the gaugino--higgsino mixing is significantly suppressed. 
The approximate plateau of $S_{\tilde{\chi}^\pm_1}$ in each frame for
$M_2$ ($|\mu|$) approaching to 150 GeV from below is due to the enhanced
gaugino-chargino mixing for degenerate $M_2$ and $|\mu|$ while with the
slowly varying scalar function $F_{sf}(\tau_{\tilde{\chi}^\pm_1})$.
In contrast, the pseudoscalar function $F_{pf}(\tau_{\tilde{\chi}^\pm_1})$ 
sharply decreases as the lighter chargino mass increases, i.e.
$\tau_{\tilde{\chi}^\pm_1}$ decreases. 
Finally, the lower--right frame in
Fig.\ref{fig:fig2} shows the dependence of the CP--even and CP--odd
amplitudes on the phase $\Phi_\mu$ for the parameter set (\ref{eq:MCPX}).
The CP--odd amplitude $P_{\tilde{\chi}^\pm_1}$ is proportional to 
$\sin\Phi_\mu$ apart from a function of $\cos\Phi_\mu$ and it
reaches its maximal value of about 0.3 around $\Phi_\mu=80^{\rm o}$. But,
the CP--even amplitude $S_{\tilde{\chi}^\pm_1}$ is an increasing function of  
$\cos\Phi_\mu$, but not of $\sin\Phi_\mu$. Again, we note that
the amplitudes $S_{\tilde{\chi}^\pm_2}$ and $P_{\tilde{\chi}^\pm_2}$ from
the heavier chargino contribution can be read off from the amplitudes 
$S_{\tilde{\chi}^\pm_1}$ and $P_{\tilde{\chi}^\pm_1}$ by exploiting
the sum rules (\ref{eq:sum rules}) and the chargino masses.\s

\vskip 0.3cm

\begin{figure}[htbp]
 \hspace*{0.0cm}
 \centerline{\psfig{figure=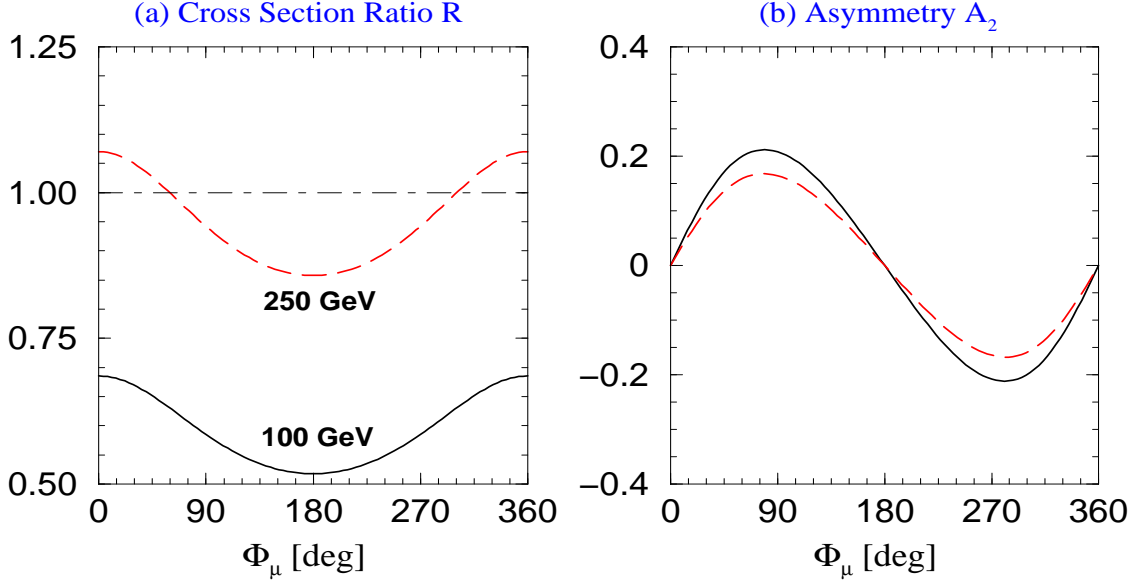,height=8cm,width=15cm}}
 \vspace*{-0.4cm}
 \caption{\it (a) the ratio $R$ of the cross section $\hat{\sigma}_0$
          to the SM prediction 150 fb and (b) the CP--odd asymmetry 
	  ${\cal A}_2$ as a function of the phase $\Phi_\mu$ for 
	  $m_{\tilde{t}_R}=100$ GeV (solid lines) and 
	  $m_{\tilde{t}_R}=250$ GeV (dashed lines), respectively. 
	  The other relevant SUSY parameters except the varied
	  parameter in each figure are given as in Eq.(\ref{eq:MCPX}).}
\label{fig:fig3}
\end{figure}

The chargino and right--handed stop contributions to the $h\gamma\gamma$
coupling affect both the Higgs production cross section and all the 
asymmetries. Nevertheless, we find that the deviation of the CP--even 
asymmetry ${\cal A}_3$ from the unity is very tiny, maximally $\sim 2$ \% for 
the parameter set (\ref{eq:MCPX}). So, we focus simply on the production 
cross section $\hat{\sigma}_0$ in Eq.(\ref{eq:cross section}) and 
the CP--odd asymmetry ${\cal A}_2$ for our numerical demonstration. 
Then, for convenience we introduce the ratio $R$ of the cross section 
$\hat{\sigma}_0$ to the SM prediction of 150 fb for $m_h=115$ GeV.
Figure \ref{fig:fig3} shows the dependence of the ratio $R$ and the 
CP--odd asymmetry ${\cal A}_2$ on the CP phase $\Phi_\mu$ for the parameter 
set (\ref{eq:MCPX}); the solid (dashed) line in each frame is for 
$m_{\tilde{t}_R}=100$ ($250$) GeV. The right--handed 
top squark contribution destructively interferes with the SM contributions, 
reducing the production cross section significantly for small 
$m_{\tilde{t}_R}$ \cite{DDHI}; about 40\,\% for $m_{\tilde{t}_R}=100$ GeV 
but 5 \% for $m_{\tilde{t}_R}=250$ GeV. On the other hand, the chargino 
contributions to the cross section are maximal in the CP--invariant case, 
about 10 \%,  but they are significantly suppressed, i.e. almost
vanishing for the maximal CP--violating case. (There may exist
a perfect cancellation between the chargino and right--handed top
squark contributions even for the CP--invariant case $\Phi_\mu=0^{\rm o}$ for 
$m_{\tilde{t}_R}\simeq 170$ GeV, leading to $R\simeq 1$.) In contrast, 
the CP--odd asymmetry ${\cal A}_2$ is vanishing for the CP--invariant case, 
but it is maximal for nearly maximal values of the CP phase; 
$|{\cal A}_2|\simeq 0.25$ for $\Phi_\mu\simeq 80^{\rm o}$ or $280^{\rm o}$, while 
as $m_{\tilde{t}_R}$ increases, the CP--odd asymmetry decreases slightly 
in magnitude. It is, therefore, clear that the cross section ratio $R$ and the 
asymmetry ${\cal A}_2$ play a {\it complementary} r\^{o}le in 
determining the chargino as well as right--handed top 
squark contributions in the decoupling limit.\s

As the lightest Higgs boson decays mainly into $b\bar{b}$ and $c\bar{c}$
pairs as well as $\tau^+\tau^-$ pairs, it is necessary to understand the
structure of possible backgrounds from the process $\gamma\gamma\rightarrow 
b\bar{b}$ and $c\bar{c}$ and the heavy quark production through the 
resolved $\gamma$ mechanisms to our Higgs boson signal and to suppress them
as much as possible \cite{KKSZ}. However, the resolved $\gamma$ mechanisms do 
not pose severe background problems in the kinematical configurations relevant 
to asymmetry measurements, because for the laser energy only slightly higher 
than the Higgs threshold the resolved $\gamma g\rightarrow b\bar{b}$ and 
$gg\rightarrow b\bar{b}$ are strongly suppressed due to
the steeply falling gluon spectrum. Moreover, the background events from
$\gamma\gamma\rightarrow b\bar{b}$ can be rejected by demanding small values 
of the rapidities of the $b$ and $\bar{b}$ quarks as 
the background events are strongly peaked at zero polar angles, but
the signal events $\gamma\gamma\rightarrow h\rightarrow b\bar{b}$ are 
distributed isotropically in the center--of--mass frame.
On the other hand, the background events from the process
$\gamma\gamma\rightarrow c\bar{c}$ are expected to be effectively suppressed 
by almost perfect $\mu$--vertexing.\s

The continuum background processes do not affect the numerator of 
the asymmetry ${\cal A}_2$ so that the background events reduce the asymmetry 
simply by a suppression factor $1/[1+N_h/N_B]$. Here, $N_h(N_B)$ denote the 
number of surviving signal (background) events. Then, the statistical 
significance ${\cal S}_{{\cal A}_2}$ for the 
asymmetry ${\cal A}_2$ extracted with $\eta=\pm \pi/4$ and $P_T=\bar{P}_T=1$ 
are
\begin{eqnarray}
{\cal S}_{{\cal A}_2}=\frac{N_h}{\sqrt{N_h+N_B}}
       \left(\frac{\langle f_3 * f_3\rangle_\tau}{
                   \langle f_0 * f_0\rangle_\tau}\right)
       \, |{\cal A}_2|\,.
\end{eqnarray}
The polarization factor $\langle f_3*f_3\rangle_\tau/\langle 
f_0*f_0 \rangle_\tau$ can be as large as 0.8 if the laser is operated in the 
red/infrared regime \cite{KKSZ} and the number of background events could 
be reduced to be comparable to or less than that of signal events \cite{LC2}. 
In this case ($N_h\simeq N_B$), the significance ${\cal S}_{{\cal A}_2}
\!\simeq 0.12\, \sqrt{N_h}$ for $\Phi_\mu\simeq 80^{\rm o}$ 
or $280^{\rm o}$ giving $|{\cal A}_2|\simeq 0.21$ for $m_{\tilde{t}_R}=100$ GeV 
so that the maximal CP violation can be detected with about 1600 signal 
events at the $5$--$\sigma$ level. On the other hand, the statistical 
significance ${\cal S}_R$ for the deviation of the cross section due to 
the (mainly) right--handed top squark as well as chargino loops from the 
SM prediction reads
\begin{eqnarray}
{\cal S}_R=\frac{N_{h{\rm SM}}}{\sqrt{N_{h{\rm SM}}+N_B}}\, |R-1|\,.
\end{eqnarray}
If $N_B\simeq N_{h{SM}}$, the significance ${\cal S}_R\simeq 0.36\, (0.11) 
\sqrt{N_{h{\rm SM}}}$ for $\Phi_\mu=180^{\rm o}$ and $m_{\tilde{t}_R}=100\, 
(250)$ GeV so that the maximal deviation from the SM prediction can be 
detected with about 200 and 2000, respectively, at the $5$--$\sigma$ level. 
Certainly, a more detailed numerical analysis would be required to determine 
the statistical significances for the CP--odd asymmetry ${\cal A}_2$ and the
ratio $R$ more accurately, and to check their complementarity more
reliably, which is beyond the scope of the present work.\s

\section{Conclusions}
\label{sec:conclusion}

Charginos as well as right--handed top squark belong to the class of the 
lightest supersymmetric particles in most SUSY theories and
the couplings of the lightest Higgs boson to charginos are in general complex
so that the chargino as well as right--handed top squark contributions to the 
loop--induced coupling of the lightest Higgs boson to two photons may be 
sizable and cause CP violation even in the decoupling limit of large 
pseudoscalar mass $m_A$. We have introduced a specific benchmark scenario 
of CP violation consistent with both the EDM constraints and a commonly 
accepted mechanism for MSSM electroweak baryogenesis; small $\tan\beta$, 
almost degenerate $M_2$ and $|\mu|$ of about the electroweak scale, the 
phase $\Phi_\mu$ close to maximal and the light right--handed top squark. 
All other SUSY particles are assumed to be very heavy mainly for avoiding 
the strong EDM constraints, and so they are essentially decoupled from 
the theory.\s

Based on the benchmark scenario of CP violation, we have analyzed the 
CP--violating chargino and CP--preserving right--handed top squark 
contributions to the $h\gamma\gamma$ coupling through the fusion of the 
lightest Higgs boson in linearly polarized photon--photon collisions in 
the decoupling regime of Higgs sector in the MSSM and its related extensions.
Our detailed analysis has clearly shown that the fusion of the lightest Higgs
boson in two linearly polarized photon collisions can provide a significant 
opportunity for detecting the deviation of the production rate mainly due 
to the right--handed top squark contribution as well as due to the chargino 
contributions and for probing CP violation solely due to the CP--violating
chargino contributions {\it even in the decoupling limit of large pseudoscalar 
mass $m_A$}. We emphasize once more that the CP--violating phenomenon due to 
chargino contributions in the decoupling limit is a unique feature of 
CP--noninvariant SUSY theories.

\bigskip

\subsection*{Acknowledgments}

The work of SYC was supported by the Korea Research Foundation 
(KRF--2000--015--DS0009) and the work of JSL was supported by the
Japan Society for the Promotion of Science. BC and PK were 
supported in part by BK21 Haeksim program of Ministry of Education and
in part by KOSEF though CHEP at Kyungpook National University.\s


\begin{thebibliography}{99}

\bibitem{MSSM} H.P. Nilles, Phys. Rep. {\bf 110}, 1 (1984); H.E. Haber and
   G.L. Kane, Phys. Rep. {\bf 117}, 75 (1985); S.P. Martin, hep--ph/9709356;
   P. Fayet, Nucl. Phys. Proc. Suppl. {\bf 101}, 81 (2001).

\bibitem{Higgs} J.F. Gunion, H.E. Haber, G. Kane and S. Dawson,
   {\it The Higgs Hunter's Guide} (Perseus Publishing, Cambridge, M.A. 2000);
   K. Inoue, A. Kakuto, H. Komatsu and S. Takeshita, Prog. Theor. Phys.
   {\bf 68}, 927 (1982) [Err. {\bf 70}, 330 (1983)]; {\bf 71}, 413 (1984);
   R. Flores and M. Sher, Ann. Phys. (NY) {\bf 148}, 95 (1983); J.F. Gunion
   and H.E. Haber, Nucl. Phys. {\bf B272}, 1 (1986); {\bf B278}, 449 (1986)
   [Err. {\bf B402}, 567 (1993)].

\bibitem{LHC} ATLAS Collaboration, Technical Design Report, CERN--LHCC 99--14;
   CMS Collaboration, Technical Proposal, CERN--LHCC 94--38;
   J.G. Branson, D. Denegri, I. Hinchliffe, F. Gianotti, F.E. Paige and
   P. Sphicas, BNL--HET--01/33 [hep--ph/0110021].

\bibitem{LC1} N. Akasaka {\it et al.}, ``JLC design study", KEK--REPORT--97--1; 
   C. Adolpsen {\it et al.} [International Study Group Collaboration],
   ``International study group progress report on linear collider development",
   SLAC--R--559 and KEK--REPORT--2000--7 (April, 2000); ``TESLA: The
   superconducting electron positron linear collider with an integrated X--ray
   laser laboratory. Technical Design Report, Part 2: The Accelerator",
   eds. R. Brinkmann, K. Flottmann, J. Rossbach, P. Schmuser, N. Walker and
   H. Weise, DESY--01--011 (March, 2001).

\bibitem{LC2} ``TESLA: The superconducting electron positron linear collider 
   with an integrated X--ray laser laboratory. Technical Design Report, Part 3:    Physics at an $e^+e^-$ Linear Collider", eds. R.D. Heuer, D.J. Miller, 
   F. Richard and P.M. Zerwas, H. Weise, DESY--01--011 (March, 2001)
   [hep--ph/0106315]; American Linear Collider Working Group, T. Abe {\it et
   al.}, SLAC--R--570 (2001), hep--ex/0106055--58; ACFA Linear Collider
   Working Group, K. Abe {\it et al.}, KEK--REPORT--2001--11 (2001),
   hep--ph/0109166.

\bibitem{DKZ} A. Djouadi, J. Kalinowski and P.M. Zerwas, Z. Phys. C {\bf 70},
   437 (1996); {\it ibid.} {\bf 57}, 569 (1993); A. Djouadi, P. Janot,
   J. Kalinowski and P.M. Zerwas, Phys. Lett. {\bf B376}, 220 (1996);
   S.Y. Choi, M. Drees, J.S. Lee and J. Song, hep--ph/0204200.

\bibitem{DDHI} A. Djouadi, V. Driesen, W. Hollik and J.I. Illana, Eur. Phys.
   J. C {\bf 1}, 149 (1998); A. Djouadi, Phys. Lett. {\bf B435}, 101 (1998);
   M. Carena, H. Haber, H.E. Logan and S. Mrenna, Phys. Rev. D {\bf 65}, 
   055005 (2002).

\bibitem{GGH} D.L. Borden, D.A. Bauer and D.O. Caldwell, Phys. Rev. D {\bf
   48}, 4018 (1993); J.F. Gunion and H.E. Haber, Phys. Rev. D {\bf 48}, 5109
   (1993); S. S\"{o}ldner--Rembold and G. Jikia, Nucl. Instrum. Meth. A 
   {\bf 472}, 133 (2001); M. Melles, {\it ibid.} {\bf 472}, 128 (2001);
   D.M. Asner, J.B, Gronberg and J.F. Gunion, hep--ph/0110320 and references
   therein.

\bibitem{GG} B. Grzadkowski and J.F. Gunion, Phys. Lett. {\bf B294}, 361
   (1992).

\bibitem{KKSZ} M. Kr\"{a}mer, J. K\"{u}hn, M.L. Stong and P.M. Zerwas,
   Z. Phys. C {\bf 64}, 21 (1994).

\bibitem{CL1} S.Y. Choi and J.S. Lee, Phys. Rev. D {\bf 62}, 036005 (2000);
   J.S. Lee, hep--ph/0106327.

\bibitem{BCK} S. Bae, B.C. Chung and P. Ko, hep--ph/0205212.

\bibitem{GKPST} I.F. Ginzburg, G.L. Kotkin, S.L. Panfil, V.G. Serbo and 
   V.I. Telnov, Nucl. Instrum. Meth. A {\bf 219}, 5 (1984); E. Boos
   {\it et al.}, {\it ibid.} {\bf 472}, 100 (2001).

\bibitem{BAU} J.M. Cline and K. Kainulainen, Phy. Rev. Lett. {\bf 85},
   5519 (2000); M. Carena, J.M. Moreno, M. Quiros, M. Seco and C.E. Wagner,
   Nucl. Phys. {\bf B599}, 158 (2001) and refereces therein;
   H. Murayama and A. Pierce, hep--ph/0201261.

\bibitem{MCX} I. Laktineh, hep--ex/0205088; see also
   http://lepsusy.web.cern.ch/lepsusy. 

\bibitem{EDM} For detailed discussions including the constraints from 
   $^{199}$Hg, see T. Falk, K.A. Olive, M. Pospelov and R. Roiban, Nucl. Phys. 
   {\bf B560}, 3 (1999); S. Abel, S. Khalil and O. Lebedev, Nucl. Phys. 
   {\bf B606}, 151 (2000) and references therein.

\bibitem{TWO1} D. Chang, W.-Y. Keung and A. Pilaftsis, Phys. Rev. Lett. 
   {\bf 82}, 900 (1999); {\it ibid.} {\bf 83}, 3972 (1999); D. Chang, 
   W.-F. Chang and W.-Y. Keung, Phys. Lett. {\bf B478}, 239 (2000);
   A. Pilaftsis, Phys. Lett. {\bf B471}, 174 (1999).
   
\bibitem{TWO2} D. Chang, W.-F. Chang and W.-Y. Keung, hep--ph/0205084.

\bibitem{CKMZ} S.Y. Choi, A. Djouadi, M. Guchait, J. Kalinowski, 
   H.S. Song and P.M. Zerwas, Eur. Phys. J. C {\bf 14}, 535 (2000);
   S.Y. Choi, J. Kalinowski, G. Moortgat--Pick and P.M. Zerwas,
   {\it ibid.} C {\bf 22}, 563 (2001); {\bf 23}, 769 (2002).

\bibitem{AMSB} L. Randall and R. Sundrum, Nucl. Phys. {\bf B557}, 79 (1999);
   T. Gherghetta, G. Giudice and J. Wells, Nucl. Phys. {\bf B559}, 27 (1999);
   J. Feng and T. Moroi, Phys. Rev. D {\bf 61}, 095004 (2000); D.K. Ghosh,
   A. Kundu, P. Roy and S. Roy, Phys. Rev. D {\bf 64}, 115001 (2001) and 
   references therein.

\end{thebibliography}
\end{document}